\documentclass[aps,prd,nofootinbib,11pt]{revtex4}
\usepackage{amsfonts}
\usepackage{txfonts}
\usepackage{graphicx}
\usepackage{dcolumn}
\usepackage{bm}
\usepackage{amssymb}
\usepackage{latexsym}

\newcommand{\be}{\begin{equation}}
\newcommand{\ee}{\end{equation}}
\newcommand{\bq}{\begin{eqnarray}}
\newcommand{\eq}{\end{eqnarray}}

\begin{document}

\title{Interacting model of new agegraphic dark energy: Cosmological
evolution and statefinder diagnostic}

\author{Li Zhang}
\affiliation{Department of Physics, College of Sciences,
Northeastern University, Shenyang 110004, China}
\author{Jinglei Cui}
\affiliation{Department of Physics, College of Sciences,
Northeastern University, Shenyang 110004, China}
\author{Jingfei Zhang}
\affiliation{Department of Physics, College of Sciences,
Northeastern University, Shenyang 110004, China}
\author{Xin Zhang}
\affiliation{Department of Physics, College of Sciences,
Northeastern University, Shenyang 110004, China}

\begin{abstract}

The statefinder diagnosic is a useful method for distinguishing
different dark energy models. In this paper, we investigate the new
agegraphic dark energy model with interaction between dark energy
and matter component by using statefinder parameter pair  $\{r, s\}$
and study its cosmological evolution. We plot the trajectories of
the new agegraphic dark energy model for different interaction cases
in the statefinder plane. As a result, the influence of the
interaction on the evolution of the universe is shown in the
statefinder diagrams.

\end{abstract}



\maketitle

\section{Introduction}\label{sec:introduction}

Recent astronomical observations of type Ia supernovae (SNIa)
indicate that the universe is undergoing an accelerating expansion~
\cite{Riess98,Perl99}. This cosmic acceleration has also been
confirmed by other observations, such as observations of large scale
structure (LSS)~\cite{Tegmark04,Abazajian04} and measurements of the
cosmic microwave background (CMB)
anisotropy~\cite{Spergel03,Bennett03}. Nowadays the most
well-accepted idea is that a mysterious dominant component --- dark
energy --- with large enough negative pressure is responsible for
this the cosmic acceleration. Despite the fact that the cosmological
origin of dark energy remains enigmatic at present, physicists have
to face the intriguing physical problem and try to understand the
ultimate nature of dark energy. Among all theoretical models, the
preferred candidate of dark energy is the Einstein's cosmological
constant $\Lambda$. The simplest cosmological model is the so-called
$\Lambda$CDM (or LCDM) model, which consists of a mixture of the
cosmological constant $\Lambda$ and the cold dark matter (CDM). The
LCDM model provides an excellent explanation for the acceleration of
the universe and the existing observational data. However, the
cosmological constant faces two difficulties, namely, the
``fine-tuning" problem and the ``cosmic coincidence" problem. The
former asks why the cosmological constant observed today is so much
smaller than the Plank scale, while the latter asks why the energy
densities of dark energy and matter are on the same order today.
Theorists have made lots of efforts to try to resolve the
cosmological constant problem but all these efforts have turned out
to be unsuccessful.

In order to alleviate or even solve these two problems, many
dynamical dark energy models have been proposed, whose equation of
state is no longer a constant but slightly evolves with time. The
dynamical dark energy scenario is often realized by some
scalar-field mechanism which suggests that the energy with negative
pressure is provided by a scalar field evolving down a proper
potential. A lot of scalar field dark energy models have been
studied, such as quintessence, \textit{K}-essence, tachyon, phantom
and quintom etc.. Besides, some interacting models have been
discussed in many works to help understand or alleviate the
coincidence problem by considering the possible interaction between
dark energy and dark matter owing to their unknown nature. For
reviews of dark energy, see, e.g., Ref.~\cite{dark energy}.

On the other hand, in various dark energy models, the property of
dark energy is strongly model-dependent. In order to be capable of
differentiating those competing cosmological dark energy scenarios,
a sensitive diagnostic for the many dark energy models is a must.
For characterizing the expansion history of the universe, one
defines the geometric parameters $H=\dot{a}/a$ and
$q=-\ddot{a}/aH^2$, namely, the Hubble parameter and the
deceleration parameter; here $a(t)$ is the scale factor of the
Friedmann-Robertson-Walker (FRW) universe. It is clear that
$\dot{a}>0$ means that the universe is undergoing an expansion and
$\ddot{a}>0$ means the universe is experiencing an accelerated
expansion. The cosmic acceleration indicates that $q$ should be less
than zero. However the deceleration parameter on its own does not
characterize the current acceleration phase uniquely. The presence
of a fairly large degeneracy in $q$ is reflected in the fact that
rival dark energy models can give rise to the same value of $q_0$ at
the present time. Under such circumstances, a robust diagnostic of
dark energy, statefinder parameter pair $\{r(z), s(z)\}$, was
introduced by Sahni et al.~\cite{Sahni:2002fz} and Alam et
al.~\cite{Alam:2003sc}. In addition, more recently, two new
diagnostics of dark energy, $Om$ and acceleration probe $\bar{q}$,
were introduced by Sahni, Shafieloo and
Starobinsky~\cite{Sahni:2008xx}.

The statefinder probes the expansion dynamics of the universe
through high derivatives of the scale factor $\ddot a$ and $\dddot
a$ and is a natural next step beyond the Hubble parameter $H$
depending on $\dot a$ and the deceleration parameter $q$ depending
on $\ddot a$. The statefinder pair $\{r, s\}$ is defined as
\begin{equation}
    r\equiv\frac{\dddot a}{aH^3},\label{defr}
\end{equation}
\begin{equation}
   s\equiv\frac{r-1}{3(q-\frac{1}{2})} .\label{defs}
\end{equation}
It is a ``geometrical'' diagnostic, in the sense that it is
constructed from a space-time metric directly, and it is more
universal than ``physical'' variables, which depend upon properties
of physical fields describing dark energy. So, in order to see the
qualitatively different cosmological evolution behaviors of dark
energy models in degeneracy of $H_0$ and $q_0$, we can plot
statefinder parameter diagrams corresponding to these dark energy
models by theoretical calculation. As a reference the spatially flat
LCDM scenario corresponds to a fixed point $\{ r, s \} = \{1,0\} $
in this diagram. Departure of a given dark energy model from this
fixed point provides a good way of establishing the ``distance'' of
this model from the LCDM~\cite{Alam:2003sc}. On the other hand, the
statefinder can also be extracted from data coming from SuperNova
Acceleration Probe (SNAP) type
experiments~\cite{Sahni:2002fz,Alam:2003sc}. Therefore, the
statefinder diagnostic combined with future SNAP observations may
possibly be used to discriminate between different dark energy
models. In this paper, we just apply the statefinder diagnostic to
the new agegraphic dark energy (NADE) model.

We will investigate the features of the NADE model with interaction
with matter component from the statefinder view point. In
Sec.~\ref{sec:agegrapgic}, we will briefly review the NADE model and
introduce an interacting model of NADE. In Sec.~\ref{sec:evolution},
we will study the cosmological evolution of the interacting NADE
model. In Sec.~\ref{sec:statefinder}, we will apply the statefinder
diagnostic to the interacting NADE model. In the last section we
will give conclusions.

\section{An interacting model of new agegraphic dark energy} \label{sec:agegrapgic}

First, let us review the NADE model. So far, we cannot confirm if
dark energy imitates as a cosmological constant or a dynamical
field. Generally, theorists believe that we cannot entirely
understand the nature of dark energy before a complete theory of
quantum gravity is established~\cite{Witten:2000zk}. However, we
still can make some efforts to probe the properties of dark energy
according to some principle of quantum gravity. The holographic dark
energy model \cite{Li:2004rb} and the agegraphic dark energy
model~\cite{rgcai} are examples, possessing some significant
features of quantum gravity. The former stems from the holographic
principle and the latter is constructed in light of the
K$\acute{a}$rolyh$\acute{a}$zy relation \cite{uncertainty} and
corresponding energy fluctuations of space-time. In this paper, we
just fucus on the agegraphic dark energy model.

In general relativity, one can measure the space-time without
 any limit to accuracy. However, in the quantum mechanics, the
 well-known Heisenberg uncertainty relation puts a limit of
 accuracy in these measurements. Following the line of quantum
 fluctuations of spacetime, K\'{a}rolyh\'{a}zy and collaborators~\cite{uncertainty} made an interesting
 observation concerning the distance measurement for Minkowski
 spacetime through a light-clock Gedanken experiment; namely,
 the distance $t$ in Minkowski space-time cannot be known to a better
 accuracy than
 \begin{equation}
\delta t=\lambda t_p^{2/3}t^{1/3}~,\label{eq1}
\end{equation}
 where $\lambda$ is a dimensionless constant of order unity. We use
 the units $\hbar=c=k_B=1$ throughout this paper. Thus, one can use
 the terms like length and time interchangeably, whereas
 $l_p=t_p=1/m_p$ with $l_p$, $t_p$ and $m_p$ being the reduced
 Planck length, time and mass, respectively.

The K\'{a}rolyh\'{a}zy relation~(\ref{eq1}) together with the
 time-energy uncertainty relation enables one to estimate a quantum
 energy density of the metric fluctuations of Minkowski
 space-time. Following Refs.~\cite{r2,r3}, with respect to
 Eq.~(\ref{eq1}) a length scale $t$ can be known with a maximum
 precision $\delta t$, determining thereby a minimal detectable cell
 $\delta t^3\sim t_p^2 t$ over a spatial region $t^3$. Such a cell
 represents a minimal detectable unit of space-time over a given
 length scale $t$. If the age of the Minkowski space-time is $t$,
 then over a spatial region with linear size $t$ (determining the
 maximal observable patch) there exists a minimal cell $\delta t^3$,
 the energy of which due to the fact that the time-energy uncertainty relation cannot
 be smaller than
\begin{equation}
 E_{\delta t^3}\sim t^{-1}~.\label{eq2}
 \end{equation}
 Therefore, the energy density of metric fluctuations of
 Minkowski space-time is given by
 \begin{equation}
 \rho_q\sim\frac{E_{\delta t^3}}{\delta t^3}\sim
 \frac{1}{t_p^2 t^2}\sim\frac{m_p^2}{t^2}.~\label{eq3}
 \end{equation}
Based on the energy density~(\ref{eq3}), the so-called agegraphic
dark
 energy model was proposed in Ref.~\cite{rgcai}. In this model, as the most natural
 choice, the time scale $t$ in Eq.~(\ref{eq3}) is chosen to
 be the age of the universe
 \begin{equation}
 T=\int_0^a\frac{da}{Ha},\label{eq4}
 \end{equation}
 where $a$ is the scale factor of our universe, and $H$
 is the Hubble parameter. Thus, the energy density of the agegraphic dark
 energy is given by~\cite{rgcai}
 \begin{equation}
 \rho_q=\frac{3n^2m_p^2}{T^2},\label{eq5}
 \end{equation}
 where the numerical factor $3n^2$ has been introduced to parameterize
 some uncertainties, such as the species of quantum fields in
 the universe, or the effect of curved space-time (since the energy
 density is derived for Minkowski space-time). Obviously,
 since the present age of the universe $T_0\sim H_0^{-1}$ (the
 subscript $0$ indicates the present value of the corresponding
 quantity), the present energy density of the
 agegraphic dark energy explicitly meets the observed value
 naturally, provided that the numerical factor $n$ is of order
 unity. In addition, by choosing the age of the universe rather than
 the future event horizon as the length measure, the drawback
 concerning causality in the holographic dark energy model
 does not exist in the agegraphic dark energy model~\cite{rgcai}.

If we consider a spatially flat FRW universe
 containing agegraphic dark energy and pressureless matter, the
 corresponding Friedmann equation reads
 \begin{equation}
 H^2=\frac{1}{3m_p^2}\left(\rho_m+\rho_q\right).\label{eq6}
 \end{equation}
 It is convenient to introduce the fractional energy densities
 $\Omega_i\equiv\rho_i/3m_p^2H^2$ for $i=m$ and $q$. From
 Eq.~(\ref{eq5}), it is easy to find
 \begin{equation}
 \Omega_q=\frac{n^2}{H^2T^2}.\label{eq8}
 \end{equation}
Obviously, $\Omega_m=1-\Omega_q$ from Eq.~(\ref{eq6}). By using
 Eqs.~(\ref{eq4})$-$(\ref{eq8}) and the energy conservation
 equation $\dot{\rho}_m+3H\rho_m=0$, we obtain the equation
 of motion for $\Omega_q$,
 \begin{equation}
 \Omega_q^\prime=\Omega_q\left(1-\Omega_q\right)
 \left(3-\frac{2}{n}\sqrt{\Omega_q}\right),\label{eq9}
 \end{equation}
 where the prime denotes the derivative with respect to
 $N\equiv\ln a$.
Evidently, from the energy conservation
 equation $\dot{\rho}_q+3H(\rho_q+p_q)=0$, as well as
 Eqs.~(\ref{eq5}) and~(\ref{eq8}), it is easy to find that the
 equation of state
 (EoS) of the agegraphic dark energy
 $w_q\equiv p_q/\rho_q$ is given by~\cite{rgcai}
 \begin{equation}
 w_q=-1+\frac{2}{3n}\sqrt{\Omega_q}.\label{eq11}
 \end{equation}

 However, there are some inner inconsistencies in this model;
 for details see Ref.~\cite{rwhao}. Therefore a new version of the
 agegraphic dark energy model was proposed to resolve the difficulties by replacing the timescale $T$ in Eq.~(\ref{eq4}) with the conformal
 time $\eta$ ~\cite{r33,r34}. This new version is often called the ``new
 agegraphic dark energy'' model. In this new version, the energy density of the agegraphic dark energy
 reads
 \begin{equation}
 \rho_q=\frac{3n^2m_p^2}{\eta^2},\label{defrho}
 \end{equation}
 where
 \begin{equation}
 \eta\equiv\int_0^t\frac{dt}{a}=\int_0^a\frac{da}{a^2H}~\label{eta}
 \end{equation}
 is the conformal age of the universe.
 The corresponding fractional energy density reads
 \begin{equation}
 \Omega_q=\frac{n^2}{H^2\eta^2}.\label{defOmega}
\end{equation}
Again we consider a flat FRW universe containing the new
 agegraphic dark energy and matter. By using
 Eqs.~(\ref{eq6}), (\ref{defrho})$-$(\ref{defOmega}) and the energy
 conservation equation $\dot{\rho}_m+3H\rho_m=0$, we find that
 the equation of motion for $\Omega_q$ is
 \begin{equation}
\Omega_q^\prime=\Omega_q\left(1-\Omega_q\right)
 \left(3-\frac{2}{n}\frac{\sqrt{\Omega_q}}{a}\right).\label{dOmega}
 \end{equation}
 From the energy conservation equation
 $\dot{\rho}_q+3H(\rho_q+p_q)=0$, as well as Eqs.~(\ref{defrho})
 and~(\ref{defOmega}), it is easy to find that the EoS of the new agegraphic dark energy is given by
 \begin{equation}
 w_q=-1+\frac{2}{3n}\frac{\sqrt{\Omega_q}}{a}~.\label{w}
 \end{equation}

The NADE model has been studied extensively; see, e.g.,
Refs.~\cite{r33,r34,nadeext}. In this paper, furthermore, we shall
extend the NADE model by including the interaction between the
agegraphic dark energy and matter. We will see that the interaction
 can significantly change the cosmological evolution. Assuming that the
 agegraphic dark energy and matter exchange
 energy through interaction term $Q$, the continuity equations become
 \begin{equation}
 \dot{\rho}_q+3H\left(\rho_q+p_q\right)=-Q,\label{inter1}
 \end{equation}
 \begin{equation}
 \dot{\rho}_{m}+3H\rho_{m}=Q,\label{inter2}
 \end{equation}
where $Q$ can be assumed as some special forms.
 For convenience, here we consider only the following particular
 interaction forms:
\begin{equation}
 Q=\left\{
 \begin{array}{ll}
3\alpha_1 H\rho_q \\ \vspace{0.75mm}
3\alpha_2 H\rho_{m}\\
3\alpha_3 H(\rho_q+\rho_{m})
 \end{array}
 \right..\label{Qform}
 \end{equation}
Differentiating Eq.~(\ref{defOmega}) with respect to $\ln{a}$ and
using Eq.~(\ref{eta}), we get
 \begin{equation}
 \Omega_q^\prime=\Omega_q\left(-2\frac{\dot{H}}{H^2}
 -\frac{2}{na}\sqrt{\Omega_q}\right).\label{GOmega}
 \end{equation}
 Differentiating Eq.~(\ref{eq6}) with respect to time $t$ and combining Eqs.~(\ref{defrho})$-$(\ref{defOmega})
 ,~(\ref{inter1}) and~(\ref{inter2}), one can easily find
 \begin{equation}
 -\frac{\dot{H}}{H^2}=\frac{3}{2}\left(1-\Omega_q\right)
 +\frac{\Omega_q^{3/2}}{na}-\frac{Q}{6m_p^2 H^3} .\label{GGOmega}
 \end{equation}
 Therefore, we obtain the equation of motion for $\Omega_q$,
\begin{equation}
 \Omega_q^\prime=\Omega_q\left[\left(1-\Omega_q\right)
 \left(3-\frac{2}{na}\sqrt{\Omega_q}\right)-Q_1\right],\label{indOmega}
 \end{equation}
 where
\begin{equation}
 Q_1\equiv\frac{Q}{3m_p^2 H^3}.\label{Q1}
 \end{equation}
 From Eqs.~(\ref{defrho}), (\ref{defOmega}) and~(\ref{inter1})~, we get the
 EoS of the interacting NADE as
\begin{equation}
 w_q=-1+\frac{2}{3na}\sqrt{\Omega_q}-Q_2,\label{inw}
\end{equation}
 where
\begin{equation}
 Q_2\equiv\frac{Q}{3H\rho_q}.\label{Q2}
\end{equation}
 It is easy to see that Eqs.~(\ref{indOmega}) and ~(\ref{inw}) reduce
 to Eqs.~(\ref{dOmega}) and ~(\ref{w}) in the case of $Q=0$~(i.e. without
interaction). Therefore, we get $\Omega_q^\prime$ and EoS of the
NADE model for the cases with and without interaction. Next, let us
look into some properties of this model.

Consider first the properties of the new agegraphic dark energy
without interaction. In the radiation-dominated epoch, $w_q=-1/3$
whereas $\Omega_q=n^2a^2$; in the matter-dominated epoch, $w_q=-2/3$
whereas $\Omega_q=n^2a^2/4$; eventually, the new agegraphic dark
energy dominates; in the late time $w_q\to -1$ when $a\to\infty$,
the new agegraphic dark energy mimics a cosmological constant. (See
Ref.~\cite{r33} for more details.) It is worth noting that this NADE
model without interaction is a single-parameter model because of its
special analytic features in the radiation-dominated and
matter-dominated epochs. Concretely, in the matter-dominated epoch,
$\Omega_q=n^2a^2/4=n^2(1+z)^{-2}/4$, where $z=a^{-1}-1$ is the
redshift. Therefore, $\Omega_q(z_{ini})=n^2(1+z_{ini})^{-2}/4$ can
be used as the initial condition to solve the differential equation
of $\Omega_q$ at any $z_{ini}$ provided that it is sufficiently deep
into the matter-dominated epoch. We choose here $z_{ini}=2000$ in
the initial condition, just as in Ref.~\cite{r34}. If $n$ is given,
we can obtain $\Omega_q$ from Eq.~(\ref{indOmega}) by using the
initial condition. Then, all other physical quantities, such as
$\Omega_m(z)=1-\Omega_q(z)$ and $w(z)$, can be obtained
correspondingly.

When interaction $Q$ is included, the situation is changed. For NADE
model without interaction ($Q=0$), the EoS $w_q$ is always larger
than $-1$ and cannot cross the phantom divide $w=-1$, see
Eq.~(\ref{w}). However, if the interaction $Q\not=0$ and $Q>0$, one
can see that $w_q$ can be smaller than $-1$ or larger than $-1$ from
Eq.~(\ref{inw}). This means that the EoS $w_q$ can possibly cross
the phantom divide in the interacting NADE model. In this case, it
should be pointed out that the initial condition
$\Omega_q(z_{ini})=n^2(1+z_{ini})^{-2}/4$ with $z_{ini}=2000$ can
also be used. In matter-dominated epoch, the contribution of dark
energy to the cosmological evolution is negligible so that the
impact of dark energy on matter can be ignored. That is to say, dark
energy cannot affect the evolution behavior of matter at early times
in spite of the existence of the interaction. So, the mentioned
initial condition $\Omega_q(z_{ini})=n^2(1+z_{ini})^{-2}/4$ with
$z_{ini}=2000$, is still proper in solving the differential equation
of $\Omega_q$ in the case of $Q\not=0$. In the next section, we will
discuss the cosmological evolution of the interacting NADE model.

\section{Cosmological evolution of the interacting NADE model}\label{sec:evolution}

For the interacting NADE model, the continuity equations for dark
energy and matter can be written as Eqs.~(\ref{inter1}) and
(\ref{inter2}), where the interaction between dark energy and matter
component is characterized by $Q$. It is convenient to define the
effective EoSs for dark energy and matter as
\begin{equation}
w_q^{(e)}=w_q+{Q\over 3H\rho_q}~,
\end{equation}
\begin{equation}
w_{m}^{(e)}=-{Q\over 3H\rho_{m}}~.
\end{equation}
According to the definition of the effective EoSs, the continuity
equations for dark energy and matter can be re-expressed in forms of
energy conservation,
\begin{equation}
\dot{\rho}_q+3H(1+w_q^{(e)})\rho_q=0~,
\end{equation}
\begin{equation}
\dot{\rho}_{m}+3H(1+w_{m}^{(e)})\rho_{m}=0~.
\end{equation}
Taking aforementioned three cases of interaction, one can obtain
\begin{equation}
 w_{m}^{(e)}=\left\{
 \begin{array}{ll}
 \frac{-\alpha_1\Omega_q}{\left(1-\Omega_q\right)} & {\rm ~for~~} Q=3\alpha_1 H\rho_q \\ \vspace{0.75mm}
 -\alpha_2\  & {\rm ~for~~} Q=3\alpha_2 H\rho_{m}\\
 -\alpha_3\left[1+\frac{\Omega_q}{\left(1-\Omega_q\right)}\right] & {\rm ~for~~} Q=3\alpha_3 H(\rho_q+\rho_{m})
 \end{array}
 \right..
 \end{equation}

Considering a spatially flat FRW universe with dark energy
$\rho_{q}$ and matter $\rho_{m}$, the Friedmann equation can be
expressed as
\begin{equation}
H(a)=H_0E(a)~,
\end{equation}
where
\begin{equation}
E(a)=\left[\frac{(1-\Omega_{q0})e^{-3\int_1^a{(1+w_{m}^{(e)})}dlna}}{1-\Omega_{q}}\right]^{1/2}~.
\end{equation}


\begin{figure}[!htbp]
\includegraphics[width=6.8cm]{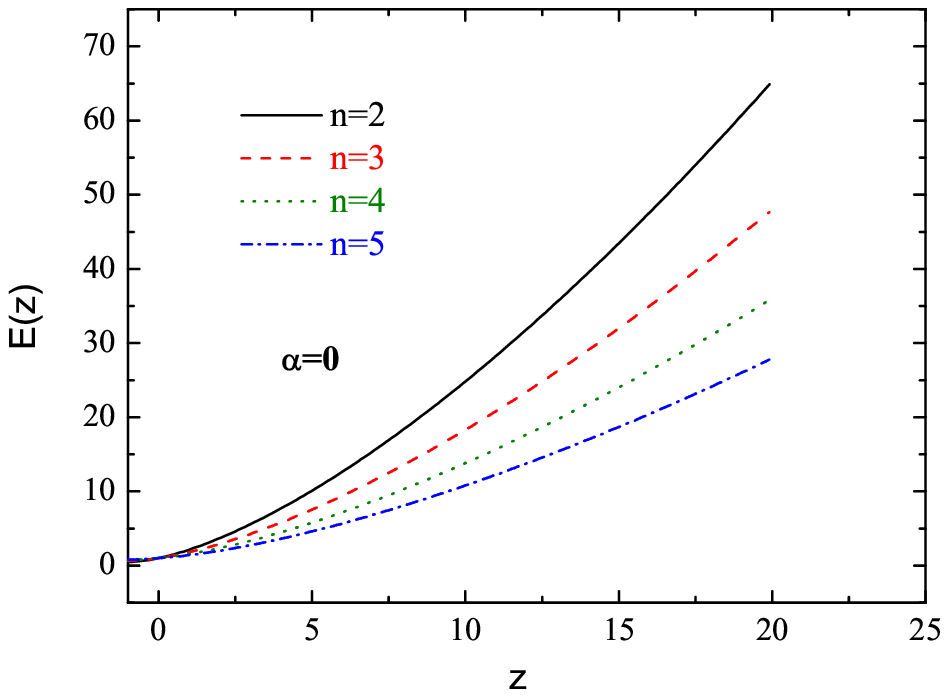}~~~
~~~~~~\includegraphics[width=6.8cm]{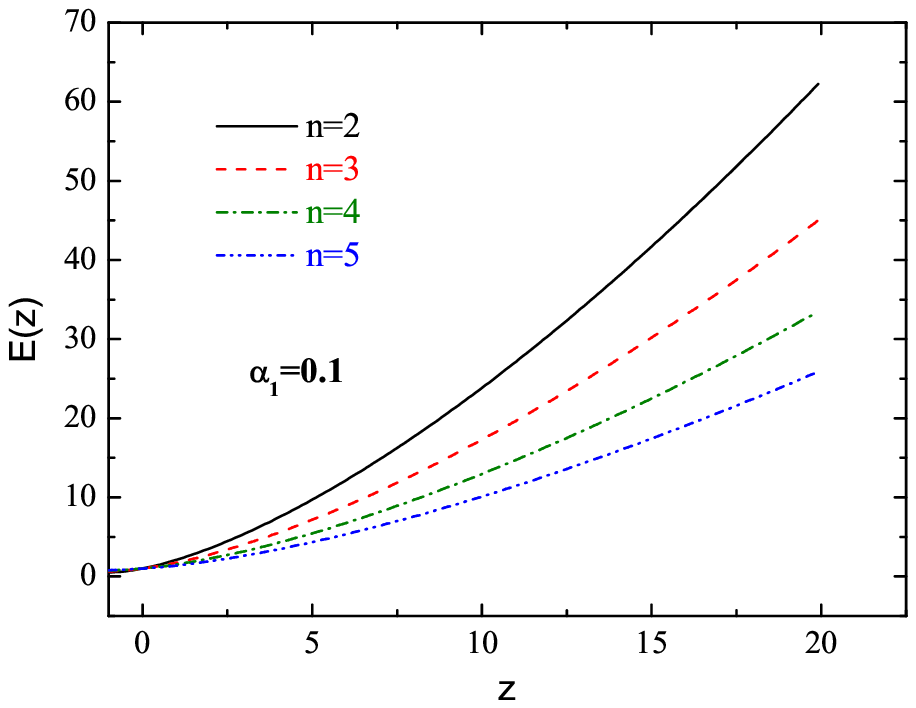}\\\vskip.3cm
\includegraphics[width=6.8cm]{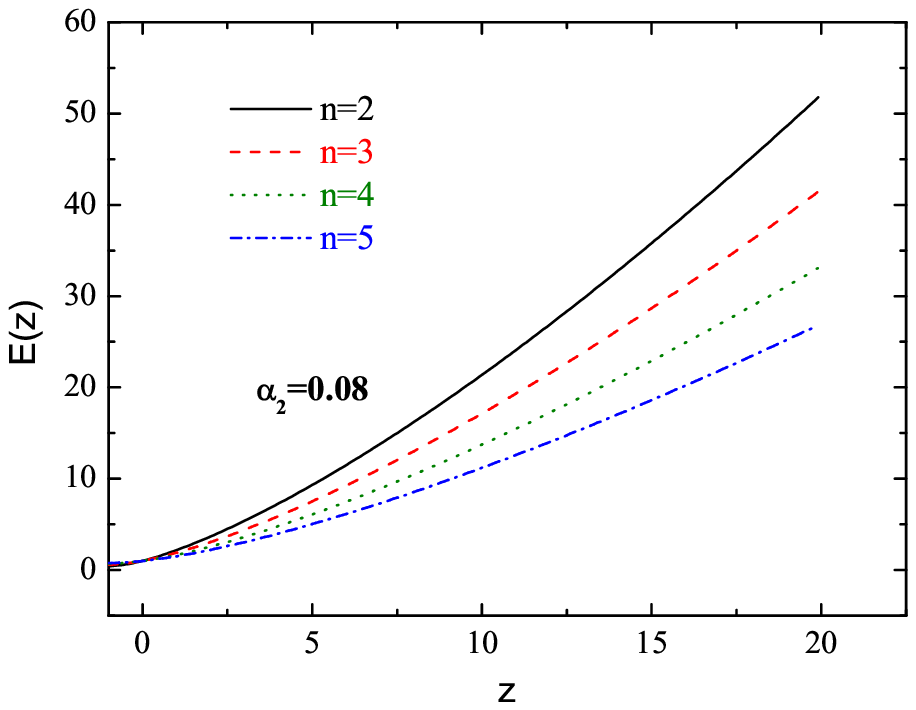}~~~
~~~~~~\includegraphics[width=6.8cm]{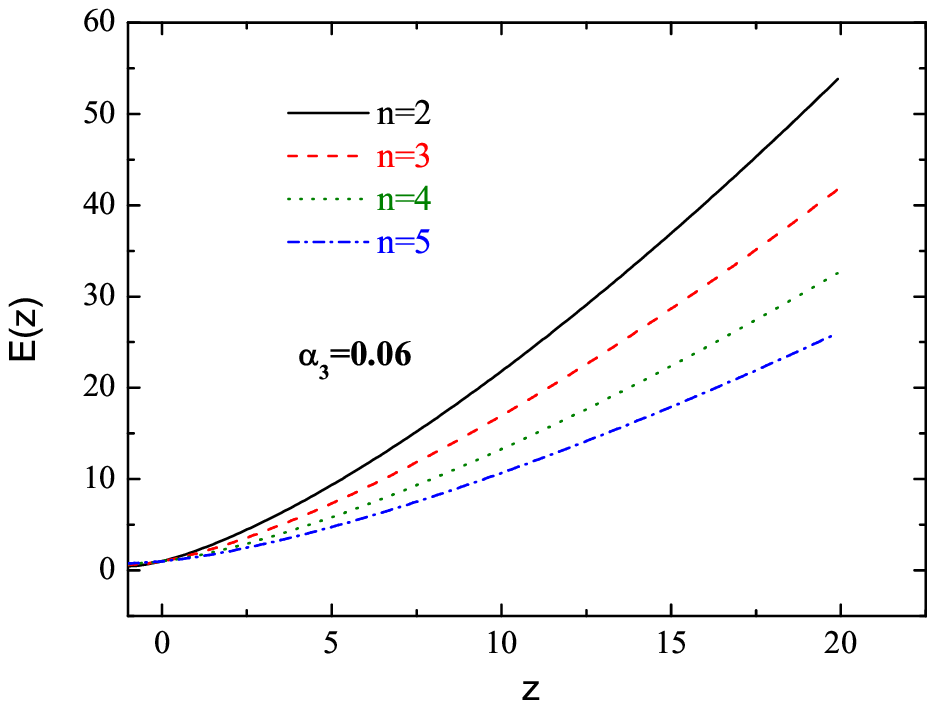} \caption{The evolution
of $E(z)$ for the interacting NADE model with the parameter $n=2$,
3, 4 and 5, respectively.}\label{fig:hn}
\end{figure}


\begin{figure}[!htbp]
  \includegraphics[width=6.8cm]{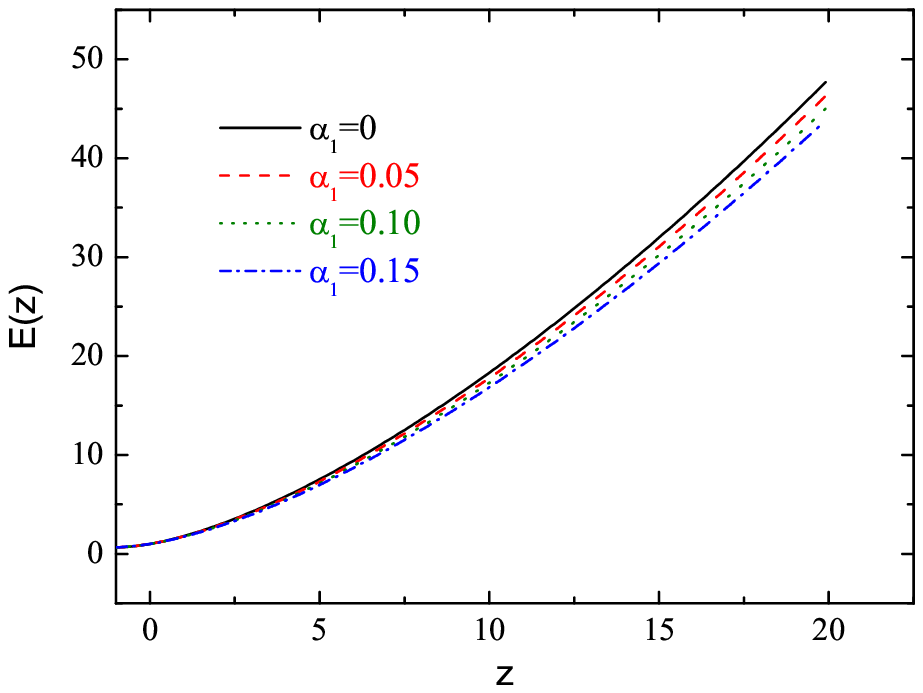}
~~~~~~\includegraphics[width=6.8cm]{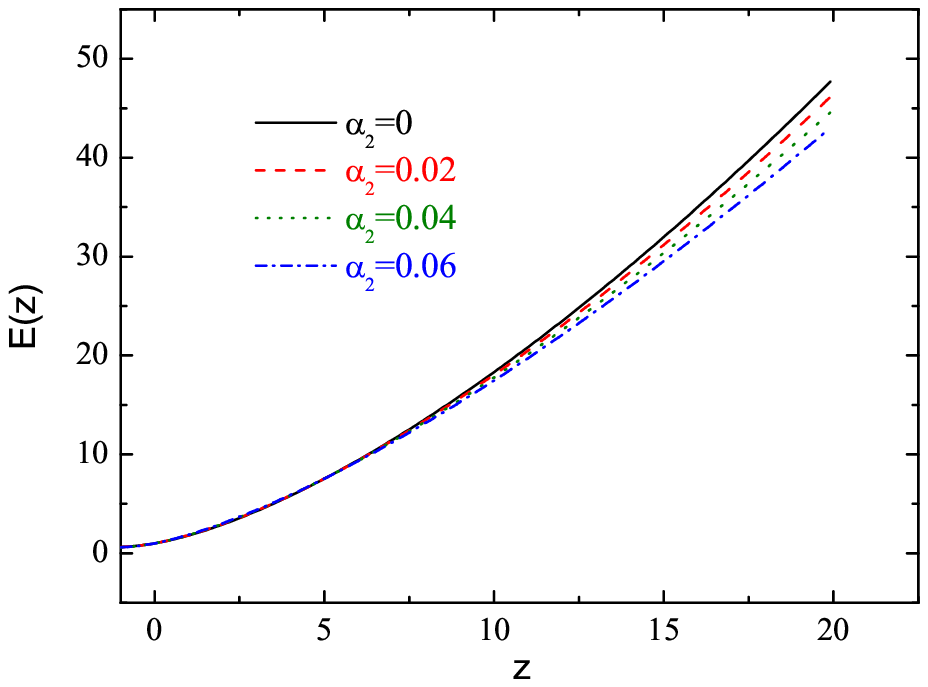}\vskip.5cm
\includegraphics[width=6.8cm]{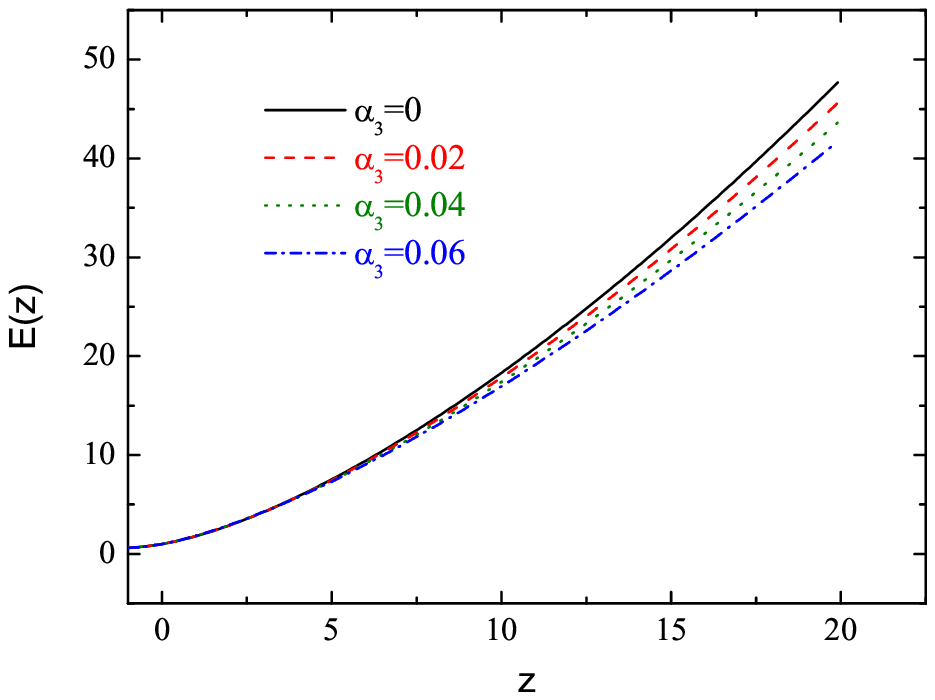}\\
\caption{The evolution of $E(z)$ for the interacting NADE model with
fixed $n$ ($n=3$) and different interaction forms.}\label{fig:ha}
\end{figure}

We plot the cosmological evolution of $E(z)$ in Figs.~\ref{fig:hn}
and~\ref{fig:ha}. First, we fix the interaction parameter $\alpha_i$
($i=1,~2,~3$, respectively) and vary the model parameter $n$. In
Fig.~\ref{fig:hn}, we show four cases, namely, the case without
interaction ($\alpha=0$), the case of $Q=3\alpha_1H\rho_q$ with
$\alpha_1=0.1$, the case of $Q=3\alpha_2H\rho_m$ with
$\alpha_2=0.08$, and the case of $Q=3\alpha_3H(\rho_q+\rho_m)$ with
$\alpha_3=0.06$. For each case, we vary the parameter $n$ and take
$n=2$, 3, 4, and 5, respectively. From this figure, we see that for
different interaction cases the cosmic evolution trends seem quite
similar, i.e., the smaller value the parameter $n$ is taken, the
bigger value the Hubble expansion rate $E(z)$ gets. Next, we fix the
model parameter $n$ (we take the case of $n=3$) and vary the
interaction parameter $\alpha_i$. In Fig.~\ref{fig:ha}, we show the
three interaction cases. Also, we see from this figure that the
cosmic evolution trends are quite similar for these three
interaction cases. The smaller the interaction parameter $\alpha_i$
is taken, the bigger the Hubble expansion rate $E(z)$ can reach.
Therefore, from the above analysis, we find that both the
parameters, $n$ and $\alpha_i$, can impact the cosmic expansion
history in the interacting NADE model.

\section{Statefinder diagnostic for the interacting NADE model}\label{sec:statefinder}

Now, let us switch to the statefinder diagnostic. In this section,
we will apply it to the interacting NADE model introduced in the
previous section. For other works on the statefinder diagnostic to
dark-energy models, see, e.g.,
Refs.~\cite{Sahni:2002fz,Alam:2003sc,statefinder}.

First, we will derive the general form of the statefinder parameters
for interacting dark energy models. The total EoS is defined as
$w_{tot}\equiv p_{tot}/\rho_{tot}
 =-1-\frac{2}{3}\frac{\dot{H}}{H^2}=-1/3+2q/3$. Also, we know that $w_{tot}=\Omega_q w_q$. So, we find
 that
 \begin{equation}
 q=\frac{1}{2}+\frac{3}{2}\Omega_q w_q.\label{qq}
 \end{equation}
From the definition of the statefinder parameter $r$ (\ref{defr}),
 it is easy to obtain
\begin{equation}
 r=\frac{\ddot{H}}{H^3}-3q-2.\label{rg}
\end{equation}
 From Eqs.~(\ref{eq6}),~(\ref{inter1}) and~(\ref{inter2}), after some
calculations, we have
\begin{equation}
 \frac{\ddot{H}}{H^3}=\frac{9}{2}+\frac{9}{2}\Omega_q w_q (w_q+2)
 -\frac{3}{2}\Omega_q w_q^\prime+\frac{3}{2}Q_1 w_q,\label{rgg}
 \end{equation}
 where
  \begin{equation}
 Q_1=\frac{Q}{3m_p^2 H^3}=\left\{
 \begin{array}{ll}
 3\alpha_1\Omega_q & {\rm ~for~~} Q=3\alpha_1 H\rho_q \\
 3\alpha_2\left(1-\Omega_q\right)  & {\rm ~for~~} Q=3\alpha_2 H\rho_{m} \\
 3\alpha_3        & {\rm ~for~~} Q=3\alpha_3 H(\rho_q+\rho_{m})
 \end{array}
 \right..
 \end{equation}
Substituting Eqs.~(\ref{qq}) and~(\ref{rgg}) into
 Eq.~(\ref{rg}), we finally obtain
\begin{equation}
 r=1+\frac{9}{2}\Omega_q w_q (1+w_q)
 -\frac{3}{2}\Omega_q w_q^\prime+\frac{3}{2}Q_1 w_q.\label{rr}
\end{equation}
 From the definition of the statefinder parameter $s$ (\ref{defs}) and Eqs.~(\ref{qq}),
 (\ref{rr}), it is easy to find that
\begin{equation}
 s=1+w_q-\frac{w_q^\prime}{3w_q}+\frac{Q_1}{3\Omega_q}.\label{ss}
 \end{equation}
From Eqs.~(\ref{indOmega})
 and~(\ref{inw}), we have
 \begin{equation}
 w_q^\prime=\frac{\sqrt{\Omega_q}}{3n}\left[\left(1-\Omega_q\right)
 \left(3-\frac{2}{na}\sqrt{\Omega_q}\right)-Q_1\right]-Q_2^\prime ,\label{indw}
 \end{equation}
where $Q_2^\prime$ represents the derivative of $Q_2$ with respect
to $\ln a$, and the form of $Q_2$ can be written as:
 \begin{equation}
 Q_2=\frac{Q}{3H\rho_q}=\left\{
 \begin{array}{ll}
 \alpha_1 & {\rm ~for~~} Q=3\alpha_1 H\rho_q \\ \vspace{0.75mm}
 \alpha_2\left(\Omega_q^{-1}-1\right) & {\rm ~for~~} Q=3\alpha_2 H\rho_{m}\\
 \alpha_3\Omega_q^{-1}   & {\rm ~for~~} Q=3\alpha_3 H(\rho_q+\rho_{m})
 \end{array}
 \right..
 \end{equation}

Now the statefinder parameters $r$ and $s$ can be theoretically
calculated for the interacting NADE model, provided that the
parameters $n$ and $\alpha_i$ are given. In what follows we shall
plot the evolution trajectories in the statefinder planes and
analyze this model from the statefinder viewpoint.

\begin{figure}[!htbp]
\includegraphics[width=6.8cm]{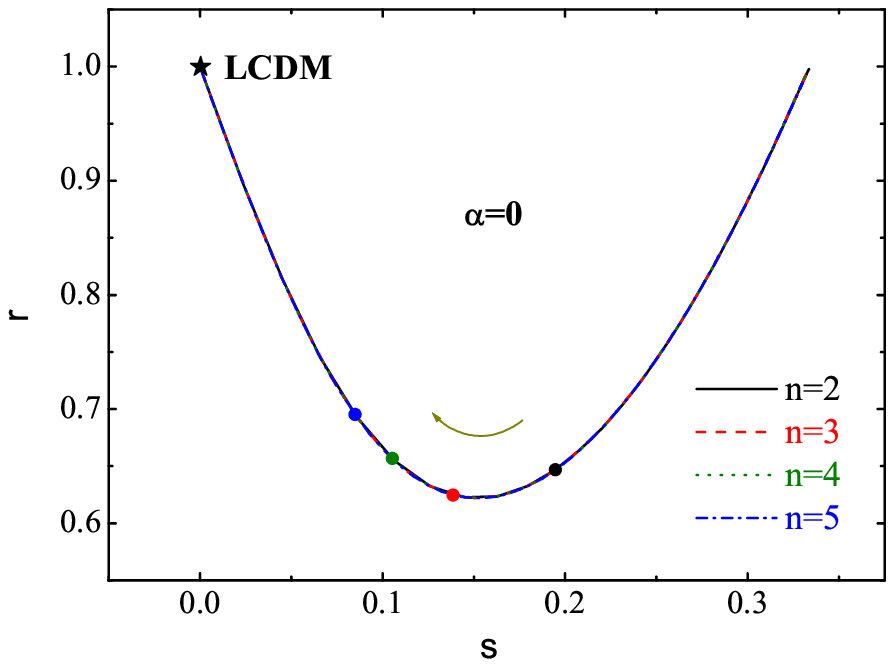}~~~
~~~~~~\includegraphics[width=6.8cm]{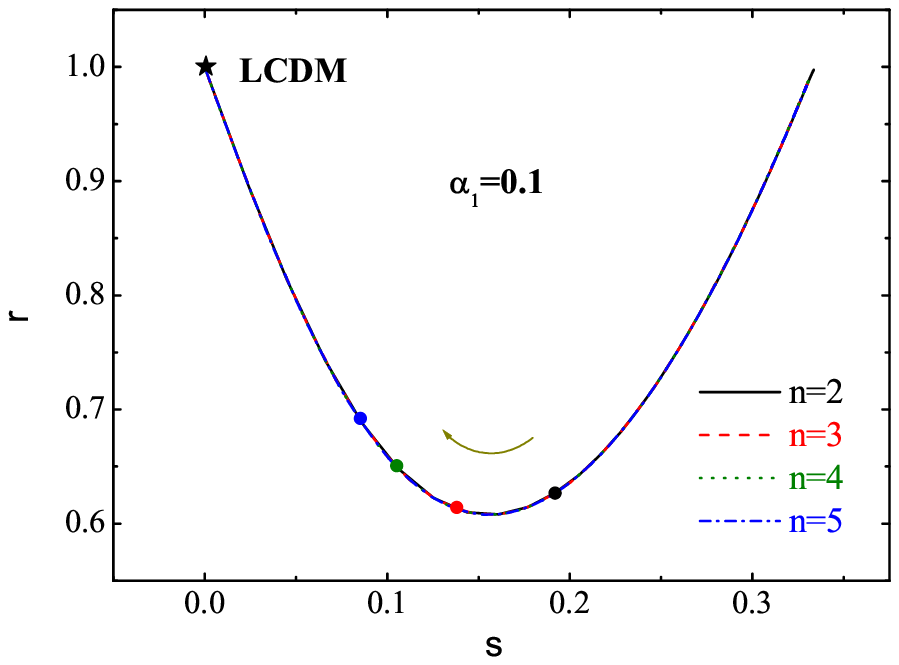}\\\vskip.3cm
\includegraphics[width=6.8cm]{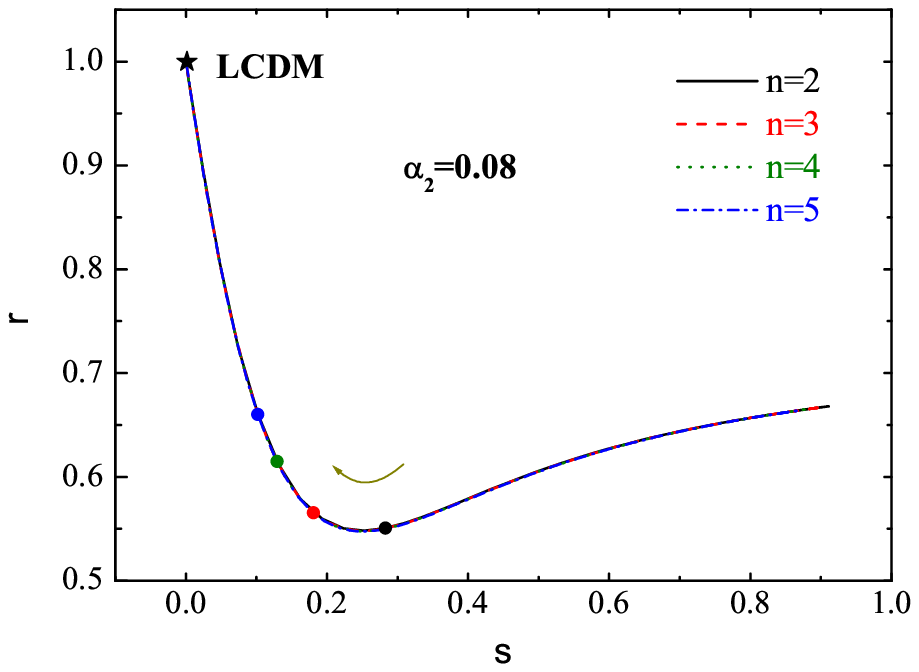}~~~
~~~~~~\includegraphics[width=6.8cm]{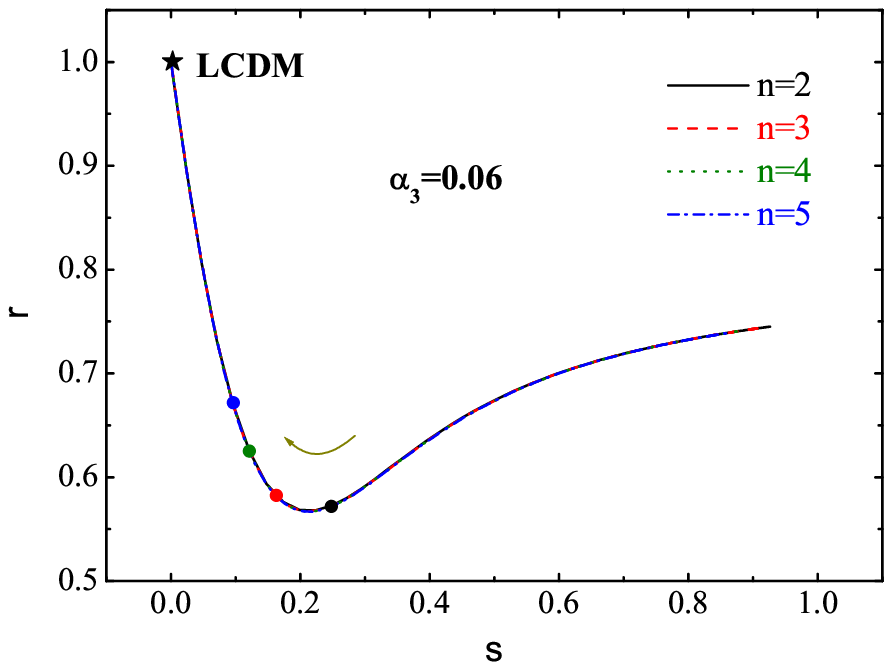} \caption{The
statefinder diagram $r(s)$ for the interacting NADE model with the
parameter $n= 2$, 3, 4 and 5, respectively. We take four cases
without and with interactions. A star denotes the LCDM fixed point
$(0, 1)$. The dots show today's values of the statefinder parameters
$(s_0, r_0)$. It is interesting to find that the curves of each
group, corresponding to different values of $n$, overlap together.
It is also rather clear that the present values of parameters,
$(s_0,r_0)$, are significantly distinguished because of the
different values of the parameter $n$, though all curves end at the
LCDM fixed point $(0,1)$.}\label{fig:srn}
\end{figure}

\begin{figure}[!htbp]
  \includegraphics[width=6.8cm]{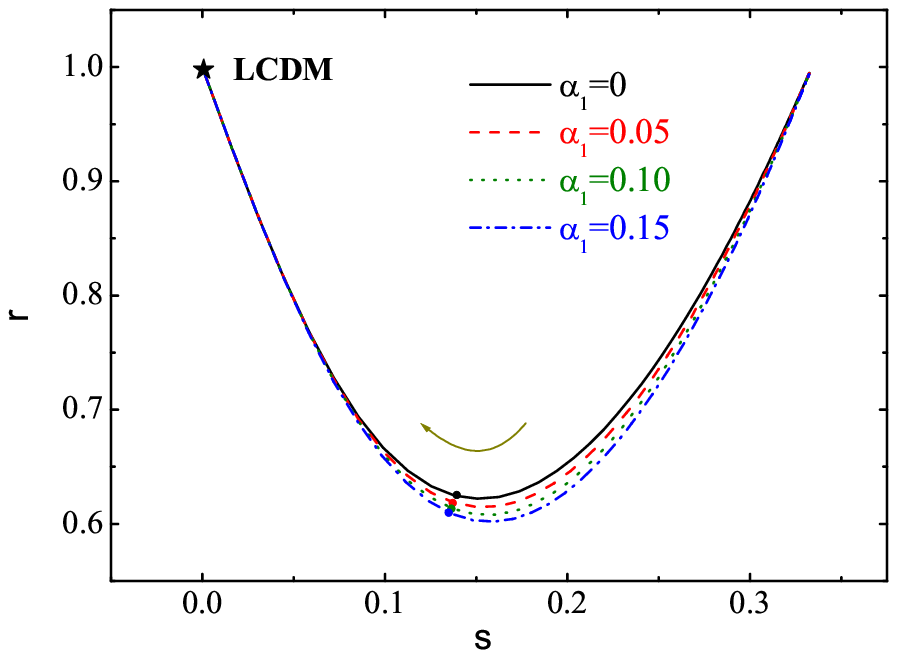}
~~~~~~\includegraphics[width=6.8cm]{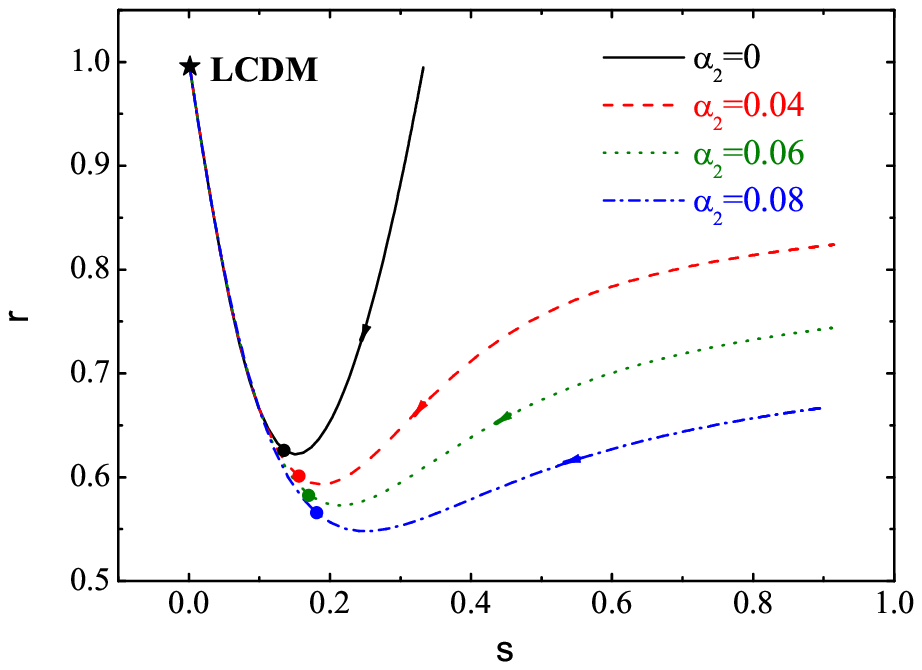}\vskip.5cm
\includegraphics[width=6.8cm]{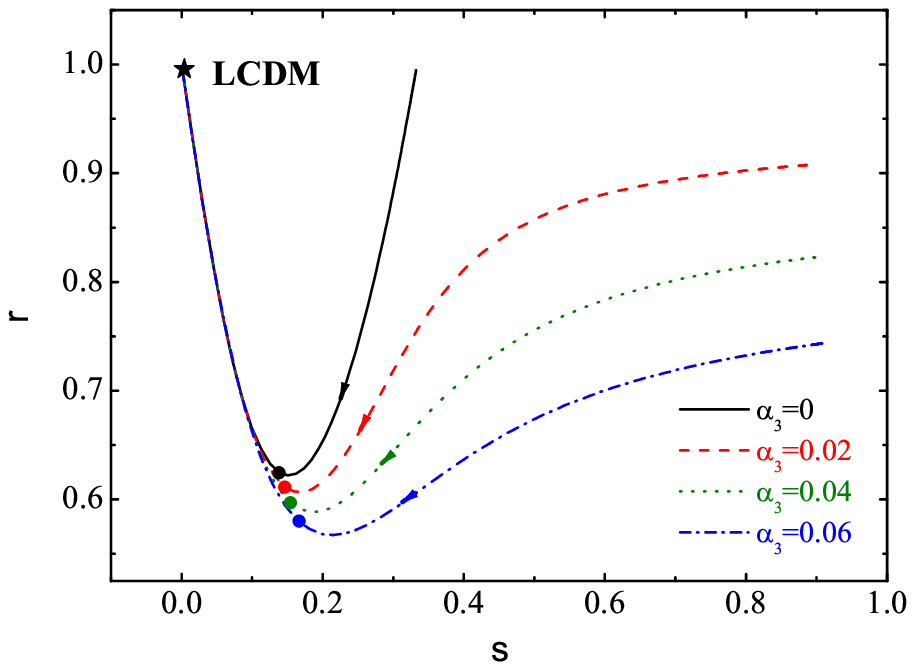}\\
\caption{The statefinder diagram $r(s)$ for the interacting NADE
model with fixed $n$ ($n=3$) and different forms of interaction. A
star denotes the LCDM fixed point $(0, 1)$. The dots show today's
values of the statefinder parameters $(s_0, r_0)$.}\label{fig:sra}
\end{figure}

We plot the statefinder diagram in the $s-r$ planes in
Figs.~\ref{fig:srn} and~\ref{fig:sra}. The case $\alpha=0$
corresponds to the case without interaction between dark energy and
matter. The arrows in the diagram denote the evolution directions of
the statefinder trajectories, and the star corresponds to
$\{r=1,s=0\}$ representing the LCDM model. In Fig.~\ref{fig:srn}, we
fix the interaction parameter $\alpha_i$ ($i=1,~2,~3$, respectively)
and vary the model parameter $n$. It is interesting to find that the
curves of each group, corresponding to different values of $n$, are
all degenerate. 
It should be mentioned that the NADE model is a single-parameter
model, i.e., only the parameter $n$ plays an important role in this
model. Figure~\ref{fig:srn} shows that the present values of
parameters $\{r, s\}$ are significantly distinguished because of the
different values of the parameter $n$, though all curves end at the
LCDM fixed point $\{r=1,s=0\}$. If the accurate information of
$\{r_0, s_0\}$ can be extracted from the future high-precision
observational data in a model-independent manner, the different
features in this model can be discriminated explicitly by
experiments, and thus one can use this method to test the NADE model
as well as other dark energy models. Hence, today's values of $\{r,
s\}$ play a significant role in the statefinder diagnosis. We thus
calculate the present values of the statefinder parameters for
different cases in the interacting NADE model and mark them on
evolution curves with dots. It can be seen that the larger model
parameter $n$ results in the shorter distance from the point $\{r_0,
s_0\}$ to the LCDM fixed point. In addition, in Fig.~\ref{fig:srn},
the first panel with $Q=0$ and the second one with $\alpha_1$ have
the similar behavior, while the situations of the third and fourth
panels are similar.


We also plotted the statefinder diagram in the $s-r$ plane for
different values of interaction parameter $\alpha_i$ with the model
parameter $n$ fixed ($n=3$); see Fig.~\ref{fig:sra}. In the left
panel of Fig.~\ref{fig:sra}, the interaction takes the form
$Q=3\alpha_1H\rho_q$ with $\alpha_1=0$, $0.05$, $0.10$ and $0.15$,
respectively. In the middle panel, the interaction
$Q=3\alpha_2H\rho_m$ with $\alpha_2=0$, $0.04$, $0.06$ and $0.08$
while in the right panel the interaction takes the form
$Q=3\alpha_3H(\rho_q+\rho_m)$ with $\alpha_3=0$, $0.02$, $0.04$,
$0.06$, respectively. With the same value of $n$, we can see that
the evolution trajectories for the case with interaction are
tremendously distinct from that of NADE model without interaction.
Moreover, the interaction of different forms will lead to different
evolutionary behavior in the statefinder parameter plane.
Concretely, just as we can see from the left panel of
Fig.~\ref{fig:sra}, when the interaction takes the form
$Q=3\alpha_1H\rho_q$, all curves evolve from the same point to the
LCDM fixed point. However, including $\rho_m$ in the interaction $Q$
(see the other two panels of Fig.~\ref{fig:sra}), the trajectories
of the interacting NADE model become very different: All curves have
the same end-point (the LCDM fixed point) but they do not begin from
the same point. We also calculate the present values of the
statefinder parameters $\{r, s\}$ for each cases and mark them on
evolutionary curves with dots in this figure. It is easy to see that
the interaction will affect the today's value of statefinder
parameter. For the interaction $Q=3\alpha_1H\rho_q$, the stronger
interaction results in the shorter distance to the LCDM fixed point.
For the cases with the interaction $Q=3\alpha_2H\rho_m$ and
$Q=3\alpha_3H(\rho_q+\rho_m)$, the bigger value of the interaction
parameter leads to the longer distance to the LCDM fixed point. From
Figs.~\ref{fig:srn} and~\ref{fig:sra}, we can learn that the
interaction between dark components makes the value of $r$ smaller
and the value of $s$ bigger, evidently. Also, obviously, the
parameter $n$ plays a crucial role in the interacting NADE model.

\section{Concluding remarks}\label{sec:conclusion}

In summary, we have studied the interacting NADE model from the
statefinder viewpoint in this paper. Since the accelerated expansion
of the universe was discovered by astronomical observations, many
cosmological models have been proposed to interpret this cosmic
acceleration. This leads to a problem of how to discriminate between
these various contenders. The statefinder diagnosis is a useful tool
for distinguishing different cosmological models by constructing the
parameters $\{r, s\}$ using the higher derivative of the scale
factor. Moreover, the value of $\{r, s\}$ of today can be viewed as
a discriminator for testing various cosmological models if it can be
extracted from precise observational data in a model-independent
way. On the other hand, although we are lacking an underlying theory
of dark energy, we still can make some efforts to probe the
properties of dark energy according to some principle of quantum
gravity. The NADE model, constructed in light of the Karolyhazy
relation and corresponding energy fluctuations of space-time, is
seen to possess some features of quantum gravity theory and provides
us with an attempt to explore the essence of dark energy. In
addition, some physicists believe that the involvement of
interaction between dark energy and dark matter leads to some
alleviation and more understanding to the coincidence problem. Thus,
it is worthwhile to investigate the interacting NADE model. We do
this by applying the statefinder parameters as a diagnostic tool and
plot the statefinder trajectories in the $s-r$ plane. We learn that
the interaction between dark energy and dark matter can
significantly affect the evolution of the universe, and the
contribution of the interaction can be diagnosed out explicitly in
this method. In addition, we show cosmological evolution of $E(z)$.
For this interacting dark energy model, the parameters $n$ and
$\alpha$ both play important roles and thus affect the cosmological
evolution. But, to determine $n$ and $\alpha$, we need more precise
data provided by future experiments. We hope that the future
high-precision observations can offer more and more accurate data to
determine these parameters precisely and consequently shed light on
the essence of dark energy.

\section*{Acknowledgements}
This work was supported by the National Natural Science Foundation
of China under Grant Nos.10705041 and 10975032.

\end{document}